\documentclass{ptephy_v1}%%%%%% to generate preprint number with ptep logo

\preprintnumber{XXXX-XXXX} %%% %%% Insert preprint number here

\usepackage{multirow}
\usepackage{charter}
\usepackage{tabularx}
\usepackage{acronym}
\usepackage{float}
\usepackage{amssymb}
\usepackage{enumerate}
\usepackage{amsthm}
\usepackage{subfig}
\usepackage{graphicx}
\usepackage{mathrsfs}
\usepackage[usenames,dvipsnames,svgnames]{xcolor}
\usepackage{soul}
\begin{document}

\title{\LARGE Optimizing Parameters of Information-Theoretic Correlation Measurement for Multi-Channel Time Series Datasets in Gravitational-Wave Detectors}

%%%% To generate auto affiliation numbers please use \author{}\affil{} command

\author[1]{Piljong Jung}
\author[1]{Sang Hoon Oh}
\author[1]{Edwin J. Son}
\affil{Gravity Research and Application Team (GReAT), National Institute for Mathematical Sciences, Daejeon, 34047, South Korea \email{johnoh@nims.re.kr}}
\author[2]{Young-Min Kim}
\affil{Department of Physics, Ulsan National Institute of Science and Technology, Ulsan, 44919, South Korea}
\author[1]{John J. Oh}

%%% To include the collaborator name... Please use the command "\collaborator"
%%% For example: \collaborator{ATLAS Collaboration}

\begin{abstract}%
Data analysis in modern science using extensive experimental and observational facilities, such as a gravitational wave detector, is essential in the search for novel scientific discoveries. Accordingly, various techniques and mathematical principles have been designed and developed to date. A recently proposed approximate correlation method based on the information theory is widely adopted in science and engineering. Although the maximal information coefficient (MIC) method remains in the phase of improving its algorithm, it is particularly beneficial in identifying the correlations of multiple noise sources in gravitational-wave detectors including non-linear effects. This study investigates various prospects for determining MIC parameters to improve the reliability of handling multi-channel time-series data, reduce high computing costs, and propose a novel method of determining optimized parameter sets for identifying noise correlations in gravitational wave data.
\end{abstract}

\subjectindex{maximal information coefficient, multi-channel data, non-linear coupling}

\maketitle
%% Acronyms
\acrodef{Gw}[GW]{Gravitational wave}
\acrodef{aligo}[aLIGO]{Advanced Laser Interferometer Gravitational-wave Observatory}
\acrodef{lasso}[LASSO]{least absolute shrinkage and selection operator}
\acresetall
\section{Introduction} \label{sec:intro}
Data analysis techniques in modern science are becoming increasingly important in achieving significant discoveries and breakthroughs in the data of scientific observations and experiments. As these observations and experiments gradually became larger, the amount of accumulated data became vast; accordingly the data processing and analysis involved in extracting meaningful information have become significant in various fields of science. Hence, this has created a novel area of data science, and its foundation includes advanced data analysis algorithms and mathematical principles, as well as the progressive development of computational resources. 

Modern experimental equipment and observation facilities are becoming more complex and precise, in which there are considerable noises to be identified and mitigated to mitigate harm to the physical interpretation from data analysis. \ac{Gw} detectors, such as the \ac{aligo} \cite{TheLIGOScientific:2014jea}, Virgo \cite{TheVirgo:2014hva}, and KAGRA \cite{Akutsu:2018axf} are complex facilities with high-precision measurements. These facilities comprises highly complicated and interconnected systems affected by various electronics and devices surrounding instruments and environments. Hence, to improve the quality of data for scientific purposes, it is crucial to elucidate subsystems of the detector and its overall status, as well as categorize and mitigate the noises from the systems and the environments. In particular, owing to the interference and mutual impacts between multi-channels in the complex devices, a few non-linear couplings also limit the detector of \ac{Gw}s. To date, several efforts in identifying the correlations of transient noises have been made to characterize aLIGO \ac{Gw} detectors. In particular, a long-duration coincidence between detector's range fluctuations and disturbances from auxiliary channels has been investigated via the \ac{lasso} regression \cite{Walker:2018ylg}. From this perspective, measuring the correlation between two random variable helps to identify the noise source elements associated with a particular device, and analyze significant relevance. 

Several measures have been introduced and utilized in identifying linear relevance, such as Pearson's correlation coefficient \cite{Pearson}, Spearman rank correlation coefficient \cite{Spearman}, and Kendall's $\tau$ coefficient \cite{Kendall}, as well as non-linear relationships such as mutual information (MI) \cite{LINFOOT195785, MI1,MI2}, distance correlation (dCorr) \cite{dCorr}, correlation distributed along curve \cite{Delicado2008}, and Heller-Heller-Gorfine (HHG) distance test \cite{HHG2012} between two variables. Another method for measuring non-linear correlation, {\it maximal information coefficient (MIC)}, has been proposed in Ref.  \cite{reshef_detecting_2011} with (non-)functional relationships \cite{Zhang2014, Speed1502}. MIC designs an approach to detect non-linear association, considering the maximal MI values defined on the $a \times b$ grid in a two-variable data plane. Hence, MIC explores every possible $a\times b$ grid up to maximal bin resolutions, and selects the maximal value among the computed MIC values on grids.  The association between two datasets can be visualized when the scattered plot is depicted on the $X-Y$ plane. Reshef {\it et al.} \cite{JMLR:v17:15-308} also proposed an empirical estimator of MIC (MICe) that avoids any heuristic approach to maximize the value for all possible resolutions. Let $D$ be a set of ${\mathfrak N}$ ordered pairs, and the total grid size is restricted by $B({\mathfrak N})={\mathfrak N}^{\alpha}$ and $c$. Note that $\alpha$ is a dimensionless parameter that controls the size of grids with $0 < \alpha <1$, and $c$ is a controlling parameter for the coarseness of the discrete grid-maximization search. Then, MICe is defined as:
\begin{equation} \label{eq:mice}
MIC^{e}(X, Y, \alpha, c) = \max_{ab<B({\mathfrak N})} \Big\{\frac{\max I^{[*]}(D, a, b)}{\log_{2} \min\{a,b\}} \Big\},
\end{equation}
where $I^{[*]}(D, a, b)$ indicates that MI is maximized in the set of $a \times b$ grids whose y-axis partition is the equipartition of size $b$, if $a \le b$. (Please refer to \ref{sec:appendixA} for more details on MI and MIC.)

\begin{figure}[t!]  %%%%%%%%%%%%%%%%%%%%%%%%%%%%%%%%%%%%%%%%%%%% Fig.
\begin{center}
\includegraphics[width=\textwidth]{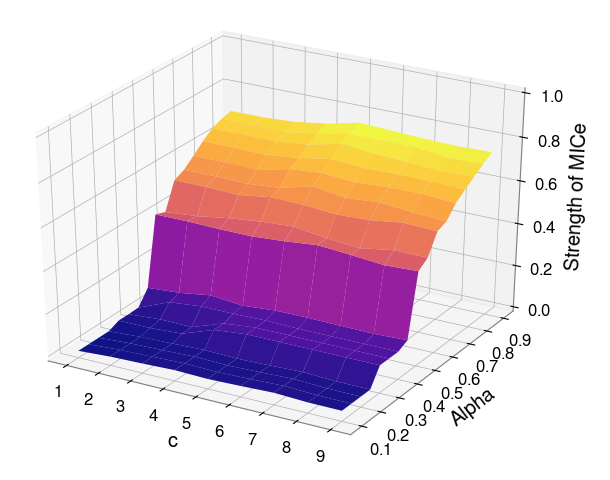}  
\caption{Plot of MICe values in Eq. (\ref{eq:mice}) for varying $\alpha$ and $c$ between two random variables.} \label{Fig.alphavsc}
\end{center}
\end{figure}

Computing MI using the grid method is simple and straightforward; however several issues need to be addressed to apply this method directly to the correlation analysis. First, there is a problem with the reliability of the estimator in choosing parameters. The value of grid-based estimators relies on choosing their parameters to establish the appropriate resolution of bins. For instance, a parameter choice can result in a non-zero value of the estimator when two variables $X$ and $Y$ are Gaussian noise sets that do not have any associations. Figure \ref{Fig.alphavsc} presents an illustrative example of how the association strength of estimators varies by selecting a set of parameters, $(\alpha, c)$. Second, we have to decide the lower bound of the data size for securing the residence of background noises, which is related to the challenge posed by the absolute value of MIC and its interpretation. For example, no significant correlation for any two random samples is expected such that its correlation coefficient should vanish. However, the estimator (MIC) is not the case for the arbitrary sample size data. This issue has been addressed, and a novel method for computing MIC combined with $\chi^2$-statistics, called ChiMIC, has been proposed \cite{chimic2016}. However, it remains insufficient to interpret the MIC results for some reliable data size and parameter sets.
Furthermore, these two issues are directly related to the problem of computational costs. With a steady increase in data size or the value of parameters, the computing costs required increase. In addition, to estimate the non-linear associations between time-series datasets, additional factors, such as data sampling rates and the type of background noises, also exert significant effects on the results. 
As pointed out in Ref. \cite{Speed1502}, MI is limited in interpreting the results; however MIC has been introduced to address its limitation by suggesting the criterion of equitability. Nevertheless, optimizing parameters to produce reliable results and interpretation remains crucial in extracting convincing information from the given data. 
We summarize the aforementioned practical issues in computing and interpreting MIC values, as expressed in the following questions:
\begin{itemize}
    \item How is the MIC value differently varied under the different types of background noises in data?
    \item What is the reliable sampling rate and data size? In addition, how do they influence the computational cost?
    \item When we handle the data from multi-channel devices with different sampling rates, does the resampling process affect MIC results? If so, what is the best way of resampling to obtain a reliable MIC score?
\end{itemize}

To answer these questions, we conduct MIC (MICe) tests with a wide range of MIC parameters and obtain a few guidelines for setting parameters when using and interpreting MIC results.

This study investigates various cases of optimizing parameters in computing MIC, and presents a methodology for parameter optimization. This research is motivated by contributions from previous studies on the formulation of MIC \cite{chimic2016, Simon2014, Justin2014}.
%and the application to KAGRA \ac{gw} data analysis \cite{Jung2021MIC}. 
In Section~\ref{sec:dataNmethods}, we discuss the process of setting up datasets for alternative/null hypotheses and the statistical power, thereby providing optimized parameters based on analytical results from the relationship between the strength and parameters presented in Section~\ref{sec:results}. Finally, we discuss our results in Section~\ref{sec:discussions}.

\section{Data and Methods}  \label{sec:dataNmethods}

We prepare datasets with some particular associations between two random variables. If noiseless random time-series $X(t)$ and $Y(t)$ can be described by $Y(X)=X$, two variables exhibit a perfect linear correlation. Similarly, we can construct a paired dataset with linear, quadratic, cubic, sinusoidal, fourth-root, circular, and stepwise associations defined by various functional relationships as presented in Table \ref{Table.FunctionTypes}. 
To generate a simulated dataset, we mix Gaussian, gamma, and Brownian noises with given datasets $X(t)$ and $Y(t)$, as well as a real instrumental noise from the \ac{Gw} detector (GWD).
%, which are plotted in Fig. \ref{Fig.NoiseTypes}. 
The GWD noise is taken from the \ac{Gw} public open data center for aLIGO. aLIGO is a 4-km long gravitational-wave detector using laser interferometry at Hanford (Washington) and Livingston (Louisiana), USA \cite{RICHABBOTT2021100658}.
It is known that the typical behavior of GWD noise is non-stationary and non-Gaussian owing to complex interconnected couplings between hundreds of thousands of instrumental and environmental noise sources.
\renewcommand\arraystretch{1.2}
\renewcommand{\tabcolsep}{5pt}
\begin{table}[htp]  %%%%%%%%%%%%%%%%%%%%%%%%%%%%%%%%%%%%%%%%%%%% Table.
\begin{tabularx}{\columnwidth}{l|l}
\hline
Functional Association Type & Function Form between $X$ and $Y$ \\ \hline
Linear   & $Y(X)=X$   \\ 
Quadratic   &    $Y(X)=4(X-\frac{1}{4})^2$    \\
Cubic    &    $Y(X)=128(X-\frac{1}{3})^3 -48(X-\frac{1}{3})^2-12(X-\frac{1}{3})$   \\
Sinusoidal: a half period   &    $Y(X)=\sin{(4\pi X)}$    \\
Sinusoidal: a quarter period  &   $Y(X)=\sin{(16\pi X)}$     \\
Fourth-root  &  $ Y(X)=X^{1/4} $    \\
Circular  &   $Y(X) = \pm \sqrt{1-(2X-1)^2}$     \\
Stepwise    &   $ Y(X)=  0  \quad \mbox{if } X \le 1/2 \quad {\rm or}~ 1 \quad \mbox{if } X > 1/2$ \\ \hline
\end{tabularx}
\caption{Datasets and functional types of random variables $X$ and $Y$ with special associations.} \label{Table.FunctionTypes}
\end{table} 

%\begin{figure}[htp]  %%%%%%%%%%%%%%%%%%%%%%%%%%%%%%%%%%%%%%%%%%%% Fig.
%\begin{center}
%\includegraphics[width=\columnwidth]{NoiseTypes.png}
%\caption{Four different noises we considered in our investigation are depicted for the sample size of $N=2048$. The GW strain data has $16384$ Hz, so we down-sampled it with the same data size for a second.} %\label{Fig.NoiseTypes}
%\end{center}
%\end{figure}

\begin{figure}[htp]  %%%%%%%%%%%%%%%%%%%%%%%%%%%%%%%%%%%%%%%%%%%% Fig.
\begin{center}
\includegraphics[width=\columnwidth]{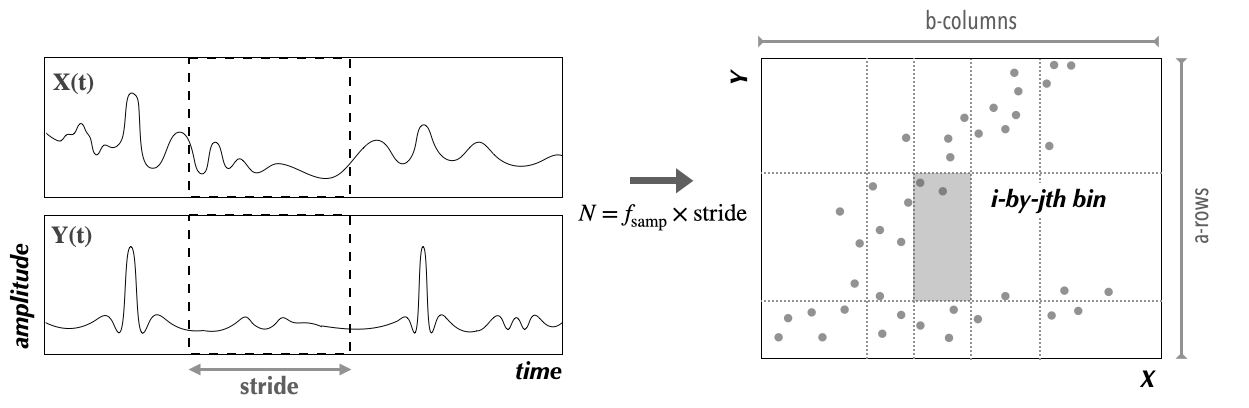}
\caption{A cartoon view of data samples from the time-series data. For a given stride, we obtain the discrete time-series data with the same sample size $N=f_{samp}\times \textrm{stride}$.} \label{Fig.time-series}
\end{center}
\end{figure}

First, we investigate the diversity of MIC values under the different types of background noises. Owing to the level of noises in various noise types (for example, Gaussian versus gamma noises), we need to define a consistent approach to interpreting compared to MIC values under the different noise backgrounds. To achieve this let us consider two time-series datasets, \textcolor{black}{$X(t)$ and $Y(t)$ which have the relation as follows, 
\begin{equation}
    \label{eq:Y(t)}
    Y = F(X) + (\textrm{RND}(n,1)-0.5)\times {\mathcal N} \times {\mathcal R},
\end{equation}
where $F$ is a functional relation shown in Table \ref{Table.FunctionTypes}, and $\mathcal R$ represents the range of $F$. And $\mathcal N$ is a relative weight of the noise amplitude and $\textrm{[RND]}$ is  randomly generated $n$ numbers in $[0,1]$ following the Ref. \cite{chimic2016}. For the time-series data, we take a time interval called stride. Since the sampling rate of the two datasets is in general not the same, we take samples with a sampling rate in a given stride, then we obtain the number of samples as $N=f_{samp}\times \textrm{stride}$. For the test of the MICe correlation, we prepare the time-series dataset for one second ($\textrm{stride}=1s $) with a different sampling rate: $f_{samp} [Hz] = \{512, 1024, 2048, 4096, 8196\}$. Therefore, the data we used has the number of data points: $N=\{512, 1024, 2048, 4096, 8196\}$. A schematic cartoon view of generating data samples from the time-series data is shown in Fig. \ref{Fig.time-series}.}

To assess an association effect of MICe for varying parameters in the different noise backgrounds, we introduce a statistical power rejecting the probability of the null hypothesis $H_0$ when the alternative hypothesis $H_1$ is true. We consider the datasets for the alternative hypothesis $H_1$, while datasets for the null hypothesis $H_0$ are randomly permuted $x$ to create the lower association data. \textcolor{black}{The set for describing $H_0$ is defined as:
\begin{eqnarray} \label{eq: null}
S_{0}(\epsilon) \equiv \big\{ {\mathcal M} = MIC^{e}\left( X_{0}(t), Y_{0}(t), \epsilon \right) | {\mathcal M} \leq D_{0}^{95{\%}}  \big\}.
\end{eqnarray}
where $D_{0}^{95\%}$ is an element of the ninety-fifth percentile in a set of $MIC^{e}$ with two random variables, $X_0$ and $Y_0$.}
Given the parameters of $\epsilon=(\alpha, c)$ in $MIC^{e}(X, Y, \epsilon)$, when we measured 500 independent observations for both hypotheses, the MICe established, satisfying the alternative hypothesis is defined as: \textcolor{black}{
\begin{equation}
S_{1}(\epsilon) \equiv \big\{ {\mathcal M} = MIC^{e}\left(X(t), Y(t), \epsilon \right) | {\mathcal M} > D_{0}^{95\%}(\epsilon) \big\}
\end{equation}
where $D_{0}^{95\%}(\epsilon) \in S_{0}(\epsilon)$.}

The set $S_{1}(\epsilon)$ containing MICe value, which are greater than an element of the set for the null hypothesis, exhibits true positive (TP). Hence, we can define a statistical power of MICe, ${\mathcal P}^{\mathrm{MICe}}$, by the ratio between the number of TP samples and the number of alternative samples $N_{1}$ for a given parameter set of $\epsilon$ as:
\begin{equation} \label{eq: power}
{\mathcal P}^{\mathrm{MICe}}(\epsilon) \equiv N[D_{1}^{5\%}(\epsilon)]/N_{1},
\end{equation}
where \textcolor{black}{$N[D_{1}^{5\%}]$ represents the number of samples that has greater MICe value than the fifth percentile of $H_1$, $D_{1}^{5\%}(\epsilon) \in S_{1}$.}

\begin{figure}[t!]  %%%%%%%%%%%%%%%%%%%%%%%%%%%%%%%%%%%%%%%%%%%% Fig.
\includegraphics[width=\columnwidth]{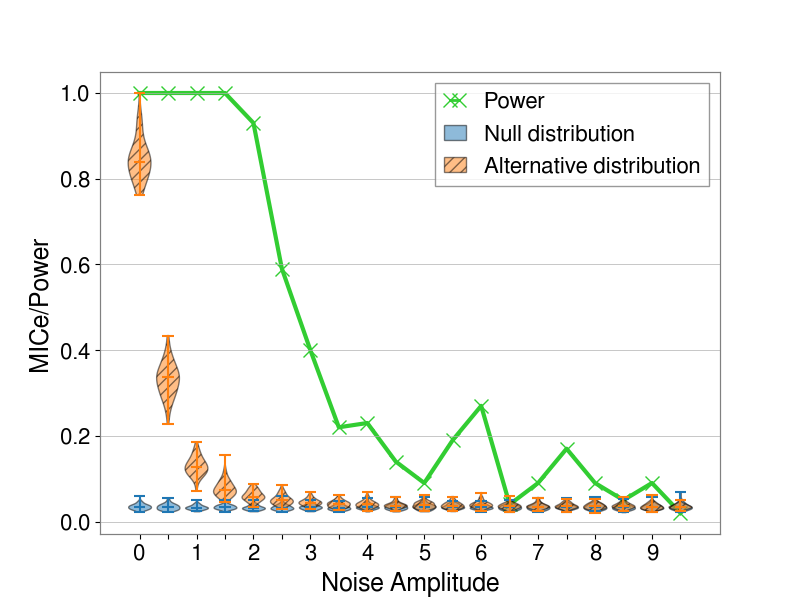}
\caption{Plot depicting the statistical power of MICe in the noise amplitude in Eq. (\ref{eq:Y(t)}) and the distribution densities of each hypothesis. \textcolor{black}{For example, we draw the power of MICe for the case of the sinusoidal association with a quarter period in Table \ref{Table.FunctionTypes}.}} \label{Fig.PowerExample}
\end{figure}

The statistical power in Eq. (\ref{eq: power}) indicates the possible separability between two hypotheses. An example of the statistical power is presented in Fig. \ref{Fig.PowerExample}, in which the statistical power of MICe for both hypotheses is plotted for diverse values of the noise amplitude levels. For low noise amplitude levels, it can be inferred that the statistical power of MICe seems efficient owing to the significant separation of both hypotheses. In contrast, the statistical power decreases when the noise amplitude level increases. We deduce that the statistical power is sufficiently efficient above $0.95$ at ${\mathcal N}$$\sim 2$ in this example.

\section{Experimental Tests} \label{sec:results}
\begin{figure}[t!]  %%%%%%%%%%%%%%%%%%%%%%%%%%%%%%%%%%%%%%%%%%%% Fig.
\includegraphics[width=\columnwidth]{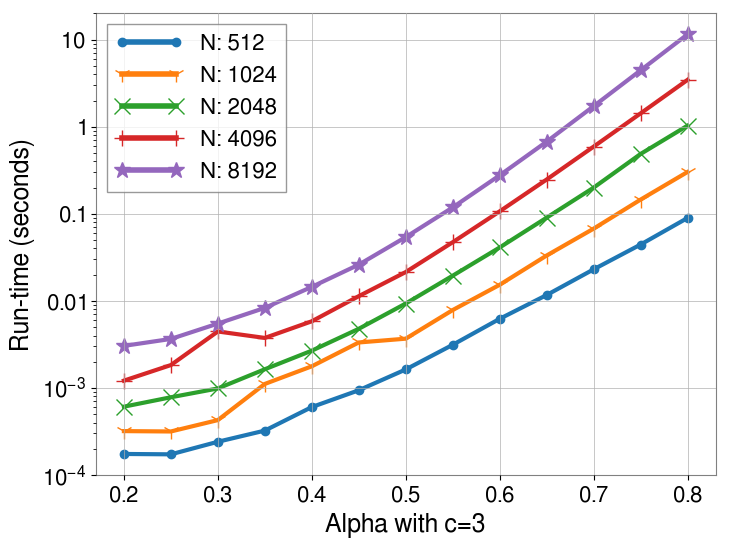}
\caption{Plot of computing runtimes \textcolor{black}{(in log-scale)} for varying $\alpha$ with a fixed $c=3$: as expected, the plot relies on the data sample size as $\alpha$ increases. \textcolor{black}{The computing runtime is the averaged value of ten computations.} } \label{Fig.ComputingCost}
\end{figure}

Based on aforementioned criterion, we estimate the statistical power of MICe for every functionally associated dataset under different background noises. To determine the effect of parameters on the power of MICe, we investigate two factors for optimizing the parameters of MICe - an {\it area under the power curve} (AUPC) and computational cost. The AUPC is defined as an area under the statistical power curve for a given parameter value presented in Fig. \ref{Fig.PowerExample}. However, we estimate by computing a sample of MICe to vary the data sample size $N$\footnote{We conducted tests in a single node of clusters with {\tt Intel(R) Xeon(R) CPU E5645@2.40GHz} and {\tt the minepy package} \cite{minepy}.} because the computing cost increases in proportion to ${\mathcal O}(c^2B(N)^{5/2}) = {\mathcal O}(c^2N^{5\alpha/2})$ as $N$ increases \cite{JMLR:v17:15-308}. As discussed in Section \ref{sec:dataNmethods}, the number of data samples improves the statistical power against the higher noise amplitude. However, the computational cost increases owing to the size of the data sample growth, as illustrated in Fig. \ref{Fig.ComputingCost}. Therefore, determining an optimal parameter set between the statistical power and the computational cost is crucial.

\subsection{Data samples and Parameter Selection}
\begin{figure*}[htp]  %%%%%%%%%%%%%%%%%%%%%%%%%%%%%%%%%%%%%%%%%%%% Fig.
\includegraphics[width=\textwidth]{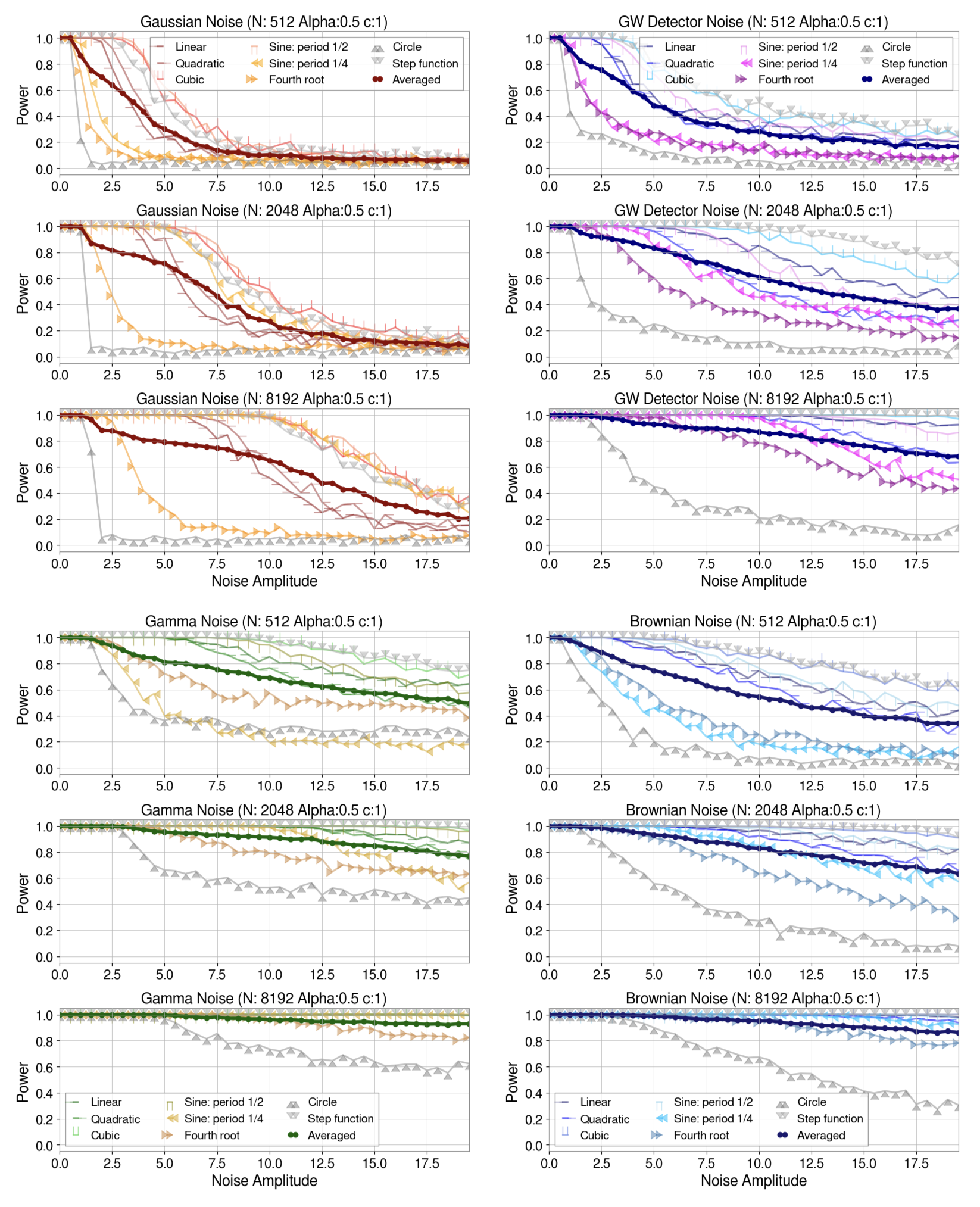}
\caption{Plots of the statistical power of MICe for eight functional associations and the averaged curve under each noise background and given parameters $(\alpha, c)=(0.5,1)$. The results are comprehensively presented in \cite{zenodo}.} \label{Fig.Powers}
\end{figure*}

Here, we first investigate the statistical power in Eq. (\ref{eq: power}) versus the noise amplitude for different numbers of data samples and different background noises. As presented in Fig. \ref{Fig.Powers}, each plot for $N=512,2048$, and $8192$ describes the statistical power as the noise amplitude increases for different background noises of the given parameters $\epsilon=(\alpha,~c)=(0.5,~1)$. In addition, the averaged curve of the statistical power for MICe is plotted with a bold marker in each figure.\footnote{Please refer to Ref. \cite{zenodo} for the whole results for $N=512,1024,2048,4096,8192$.}

For low noise amplitude levels, the power seems sufficiently effective regardless of background noises; however, the power decreases with the different levels of the efficiency as the noise amplitude level increases. When the sample size becomes large, the statistical power also remains efficient as the noise amplitude level increases, as illustrated in the plots for $N=8192$ in each plot. Except for the cases of the fourth-root and circular relationships, every association has a constantly efficient power for larger $N$ cases. In particular, it is observed that the circular association between two variables seems to be significantly sensitive to noises. Consequently sufficient data samples are required to maintain a reliable statistical power of MICe. If we consider the averaged curve of the statistical power for MICe, we may select ${\mathcal N}$$\sim 2$ for the most efficient power of MICe. However, we cannot apply this in a real case because we would be unable to elucidate the accurate association and functional relationships with the noisy datasets.

\begin{figure*}[htp]  %%%%%%%%%%%%%%%%%%%%%%%%%%%%%%%%%%%%%%%%%%%% Fig.
\includegraphics[width=\textwidth]{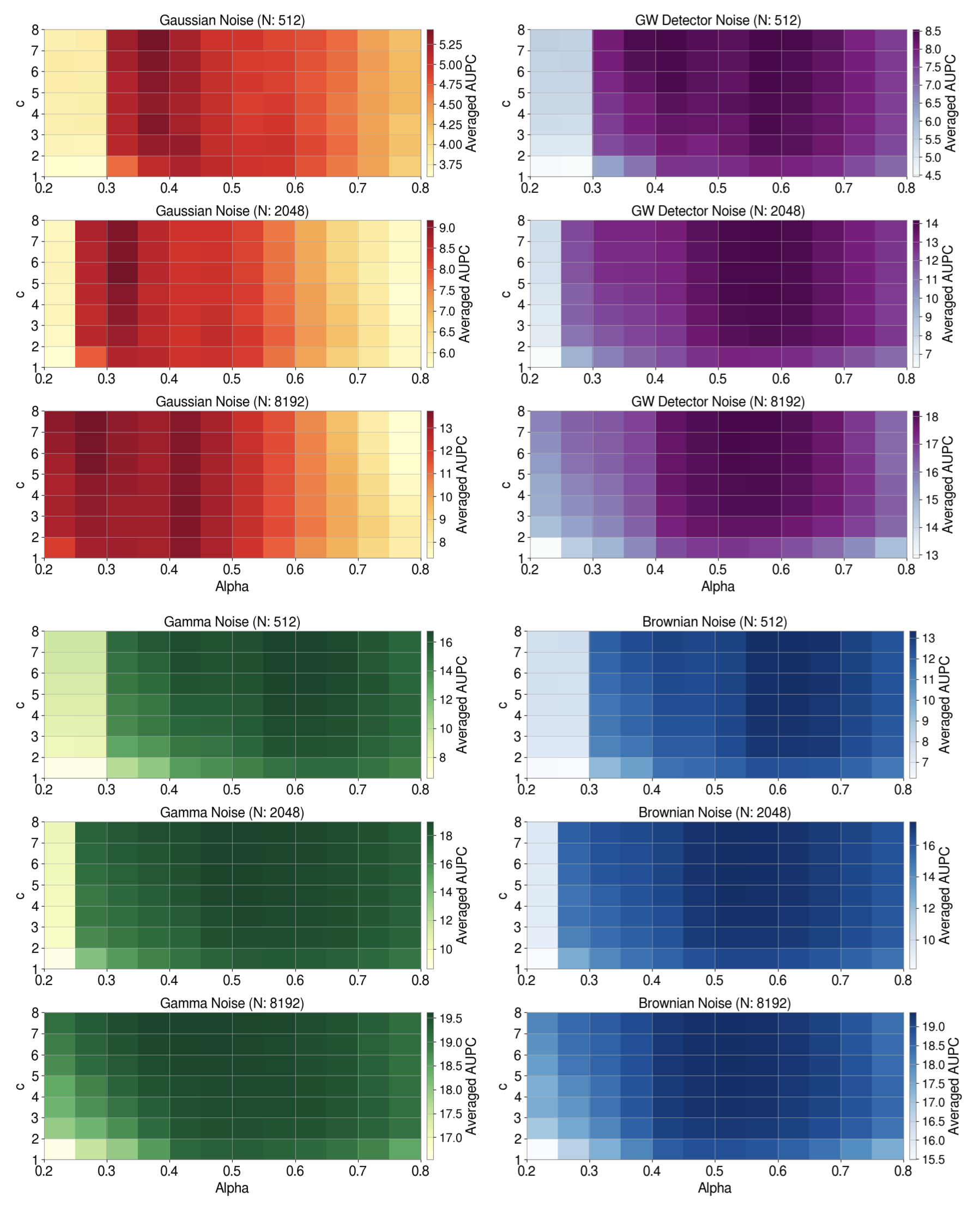}
\caption{Heat maps of the averaged AUPC under each noise background for varying $\alpha$ and $c$ when $N=512$, $2048$, and $8192$. The results are comprehensively presented in \cite{zenodo}.} \label{Fig.AUCpowers}
\end{figure*}

\begin{figure*}[htp]  %%%%%%%%%%%%%%%%%%%%%%%%%%%%%%%%%%%%%%%%%%%% Fig.
\includegraphics[width=\textwidth]{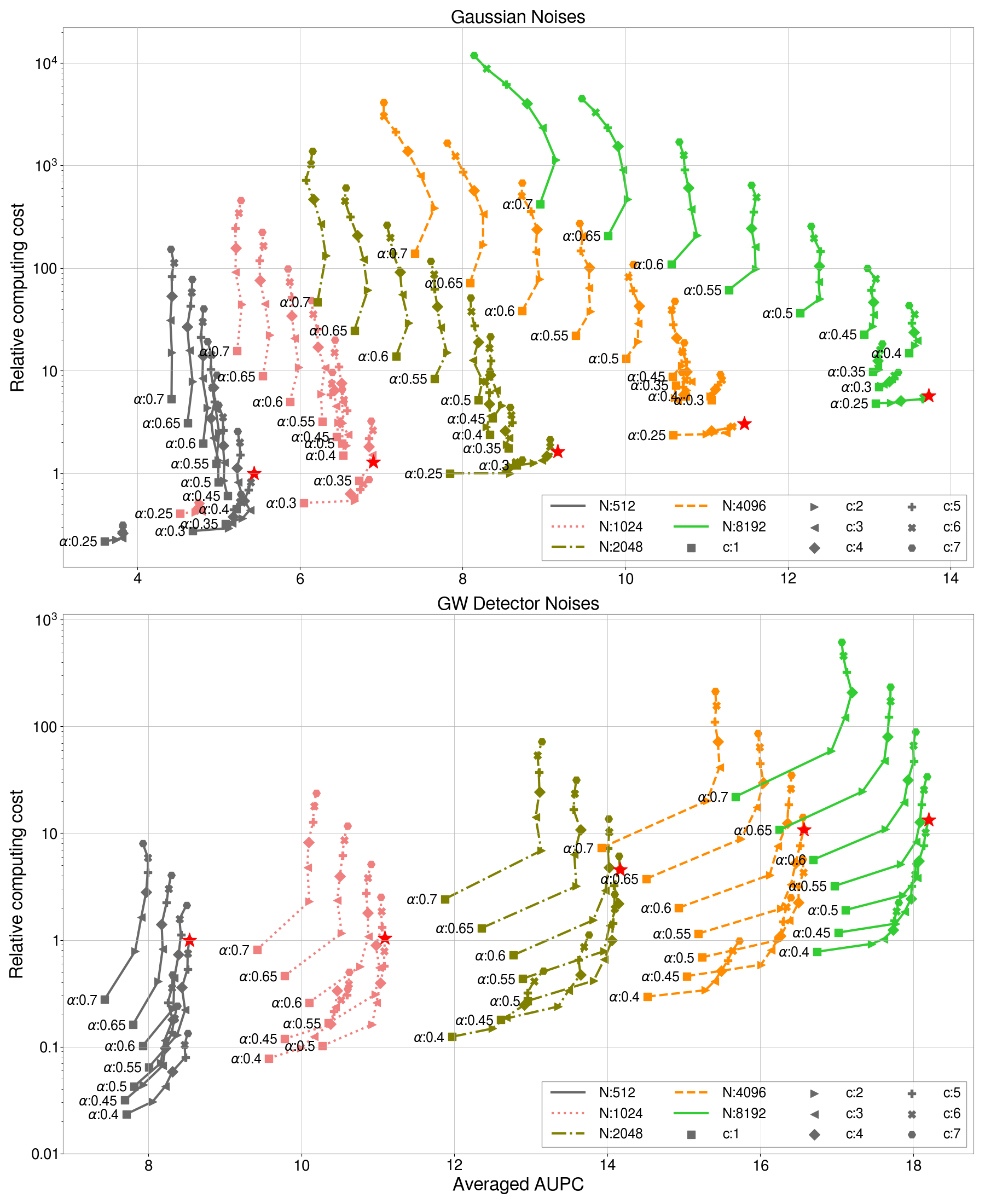}
\caption{Plots of the averaged AUPC of MICe versus the relative computational cost for varying parameters of $\epsilon=(\alpha, c)$. \textcolor{black}{Each point represents the AUPC with the corresponding computational cost for a given parameter $(\alpha, c)$.} The red asterisk mark points represent the \textcolor{black}{proposed values of parameters with the maximal averaged AUPC for relatively low computational cost of a given sample size $N$.} \textcolor{black}{Hence we conclude that the parameter value $(\alpha, c)$ of the red asterisk marks can be regarded as the maximal AUPC of MICe, \textcolor{black}{considering the relatively low} computational cost.} The analysis results are comprehensively presented in \cite{zenodo}.} \label{Fig.AUCpowersNcost}
\end{figure*}

Now, we can estimate the most optimal set of $\epsilon$ producing the highest power of MICe for different background noises. In Fig. \ref{Fig.AUCpowers}, the heatmaps of the average AUPC are illustrated relative to $N=512, 4096$, and $8192$ in the parameter space of varying $\alpha$ and $c$. Each block in this heatmap indicates the strength of the averaged AUPC with the efficient values of parameters $\alpha$ and $c$ for a given data sample and background noise. Note that each block in the heat map depicts the lower left-hand side value of $(\alpha, c)$. Therefore, by computing the averaged AUPC for varying parameters, we can select more efficient parameter sets to produce the highest statistical power of MICe. We select the parameter set of $(\alpha, c)$ and its relative computing cost for each data sample size that provides the highest averaged AUPC presented in Table \ref{tab:optparameters}. The relative computing cost is computed as a relative value based on the computational time of the Gaussian noise case for $N=512$. 

We can plot the average AUPC versus the computation cost for each noise type, and the data sample size in Fig.~\ref{Fig.AUCpowersNcost},  \textcolor{black}{which exhibits a consistent pattern of computational cost versus average AUPC that can be calculated for each number of data samples. Each point represents the average AUPC with the computational cost for a given $(\alpha, c)$. For each tail of a fixed $\alpha$ and varying $c$, we want to choose a \textcolor{black}{proposed} set of $(\alpha, c)$ with maximal AUPC, which is marked with the red asterisk.
The \textcolor{black}{proposed parameter set of $(\alpha, c)$ with the maximal AUPC} of MICe is presented in Table \ref{tab:optparameters}. }

\subsection{Resampling for Multi-channel Datasets} 
If a dataset comprises data obtained from multi-channel devices with various sampling frequencies, it is necessary to equally match the different sampling frequencies of the two channels. Here, there are three ways to do this: 1) match both equally by down-sampling the channel of the high sampling frequency (HD) 2) match both equally by up-sampling the channel of the low sampling frequency (LU) 3) match both equally by up-sampling (down-sampling) the channel of the low (high) sampling frequency into the intermediate sampling frequency (BR). Therefore, we need to investigate the effect of the resampling process on the statistical power of MICe.

\begin{figure*}[htp]  %%%%%%%%%%%%%%%%%%%%%%%%%%%%%%%%%%%%%%%%%%%% Fig.
\includegraphics[width=\textwidth]{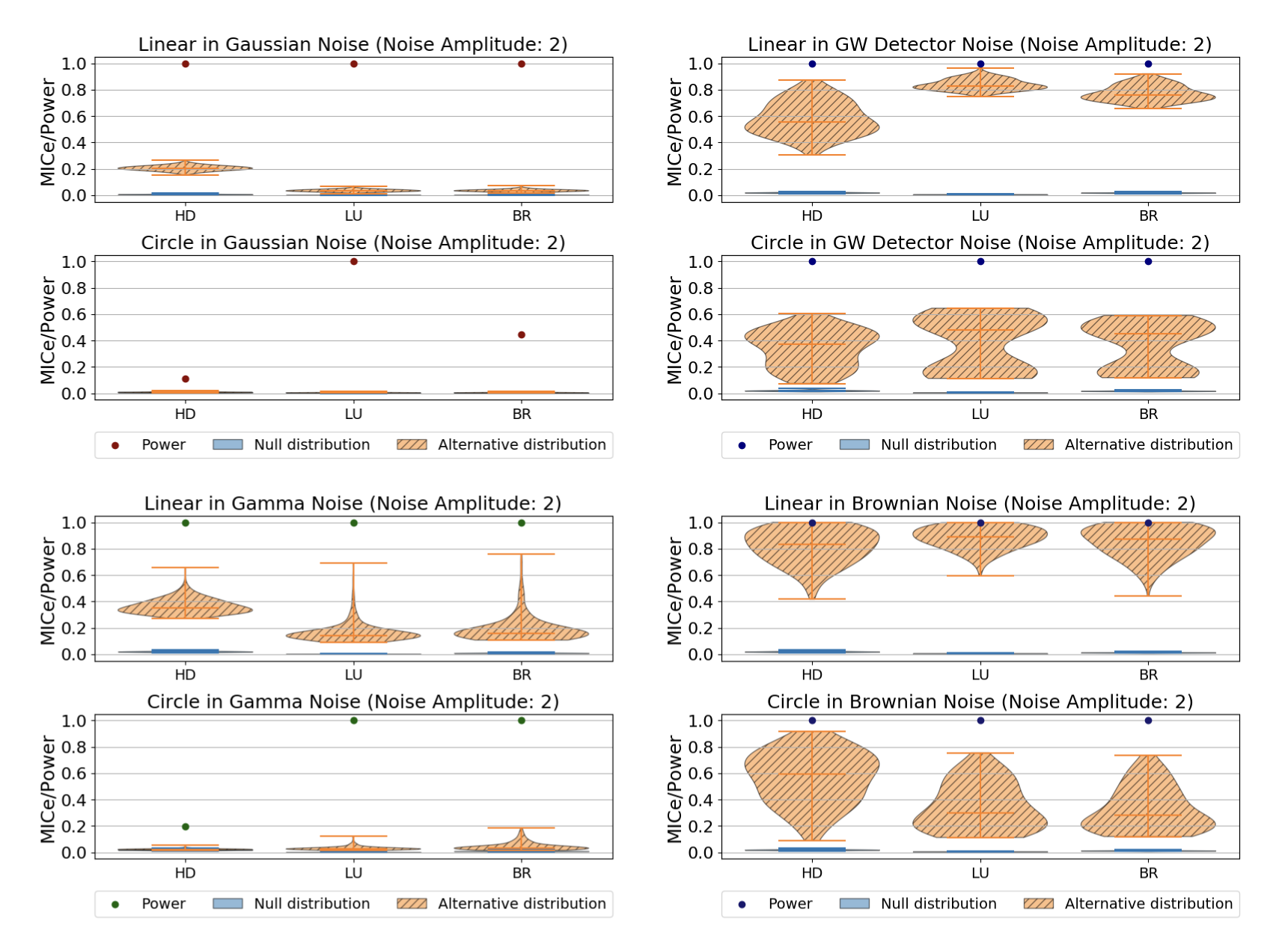}
\caption{Comparison of the statistical powers and the distribution of MICe values for resampling datasets with different scenarios under each nose background. Please refer to \cite{zenodo} for the entire analysis results.} \label{Fig.ResamplePowers}
\end{figure*}

\begin{figure}[htp]  %%%%%%%%%%%%%%%%%%%%%%%%%%%%%%%%%%%%%%%%%%%% Fig.
\includegraphics[width=\columnwidth]{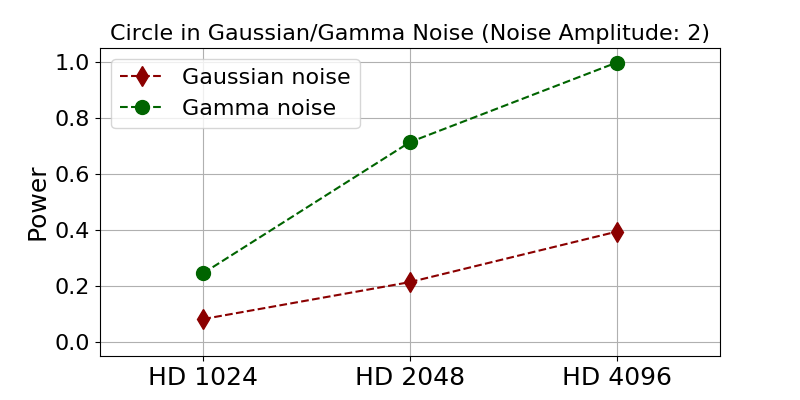}
\caption{Comparison between the statistical powers of three cases for down-samplings. The $HD 1024$, $HD 2048$, and $HD 4096$ cases represent the down-sampling to $f=1024Hz$, $2048Hz$, and $4096Hz$ from $f=8192Hz$, respectively.} \label{Fig.ResampleGaussianNGamma}
\end{figure}

To achieve this, we first prepare two datasets with different sampling frequencies, $f=8192Hz$ and $f=1024Hz$. We investigate the above three scenarios for all functional associations under different noise background levels of ${\mathcal N}=2$. The first option is to equally down-sample the data with $f=8192Hz$ to that with $f=1024Hz$ (HD) while we up-sample the data with $f=1024Hz$ to that with $f=8192Hz$ equally in the second option (LU). Finally, we resample both datasets to an intermediate sampling frequency of $f=2048Hz$ or $4096Hz$ (BR). Some analysis results for only linear and circular correlations are presented in Fig. \ref{Fig.ResamplePowers}. The plots illustrate the statistical power aspect for distinguishing null and alternative distributions when we resample datasets. Except for a specific case of circular association for Gaussian and gamma noises, the resampling effect does not affect the statistical power of MICe. This phenomenon emerges because the down-sampled data size of the circular association in the Gaussian noise case is insufficient to possess the statistical power required to distinguish null/alternative distributions. This criterion is validated for comparisons by performing another experiment with the datasets of $f=8192Hz$ and $f=1024Hz,~2048Hz$, and $4096Hz$ in Gaussian/gamma noise backgrounds in Fig. \ref{Fig.ResampleGaussianNGamma}, in which it is inferred that the statistical power increases as down-sampled datasets size increases, thus implying that it is important to obtain the data sample size after down-sampling, regardless of the resampling procedure. Therefore, we infer that if the sufficient size of data samples is guaranteed, the data resampling does not affect the statistical power of computing MICe. 

\renewcommand\arraystretch{1.4}
\renewcommand{\tabcolsep}{2pt}
\begin{table*}[!htbp]
\begin{small}
\centering
\begin{tabular}{ccccccc}
\hline
\hline
Noise Type & $N$ & $\alpha$ & $c$ & Averaged AUPC & Relative Computing Cost & Runtime (sec) \\ \hline\hline
\multirow{5}{*}{Gaussian Noise} & 512 & 0.35 & 7.0 & 5.434 & 1.000 & 7.2779$\times 10^{-4}$ $\pm$ 5.1360$\times 10^{-6}$ \\ 
 & 1024 & 0.35 & 2.0 & 6.899 & 1.286 & 9.4907$\times 10^{-4}$ $\pm$5.3110$\times 10^{-5}$\\
 & 2048 & 0.30 & 5.0 & 9.166 & 1.625 & 1.1988$\times 10^{-3}$ $\pm$ 3.5905$\times 10^{-5}$ \\
 & 4096 & 0.25 & 7.0 & 11.465 & 3.069 & 2.2643$\times 10^{-3}$ $\pm$ 5.1274$\times 10^{-5}$ \\
 & 8192 & 0.25 & 7.0 & 13.742 & 5.694 & 4.2009$\times 10^{-3}$ $\pm$ 1.5290$\times 10^{-5}$ \\\cline{1-7} 
\multirow{5}{*}{GW Detector Noise} & 512 & 0.55 & 7.0 & 8.535 & 1.000 & 1.4158$\times 10^{-2}$ $\pm$ 5.4217$\times 10^{-5}$ \\
 & 1024 & 0.50 & 7.0 & 11.092 & 1.040 & 1.4721$\times 10^{-2}$ $\pm$ 6.6200$\times 10^{-5}$ \\
 & 2048 & 0.55 & 6.0 & 14.164 & 4.561 & 6.4571$\times 10^{-2}$ $\pm$ 5.8023$\times 10^{-4}$ \\
 & 4096 & 0.55 & 6.0 & 16.566 & 10.781 & 1.5264$\times 10^{-1}$ $\pm$ 4.1442$\times 10^{-3}$ \\
 & 8192 & 0.50 & 7.0 & 18.199 & 13.330 & 1.8872$\times 10^{-1}$ $\pm$ 3.7194$\times 10^{-4}$ \\\cline{1-7} 
\multirow{5}{*}{Gamma Noise} & 512 & 0.6 & 7.0 & 16.752 & 1.000 & 2.9842$\times 10^{-2}$ $\pm$ 8.6986$\times 10^{-5}$ \\
 & 1024 & 0.50 & 7.0 & 18.234 & 0.493 & 1.4721$\times 10^{-2}$ $\pm$ 6.6200$\times 10^{-5}$ \\
 & 2048 & 0.45 & 7.0 & 18.955 & 0.531 & 1.5858$\times 10^{-2}$ $\pm$ 9.5062$\times 10^{-5}$ \\
 & 4096 & 0.40 & 7.0 & 19.346 & 0.466 & 1.3906$\times 10^{-2}$ $\pm$ 4.6712$\times 10^{-5}$ \\
 & 8192 & 0.40 & 7.0 & 19.614 & 1.069 & 3.1898$\times 10^{-2}$ $\pm$ 4.4971$\times 10^{-5}$ \\\cline{1-7} 
\multirow{5}{*}{Brownian Noise} & 512 & 0.60 & 6.0 & 13.320 & 1.000 & 2.2202$\times 10^{-2}$ $\pm$ 5.4714$\times 10^{-5}$ \\
 & 1024 & 0.55 & 7.0 & 15.736 & 1.613 & 3.5811$\times 10^{-2}$ $\pm$ 2.3834$\times 10^{-3}$ \\
 & 2048 & 0.50 & 6.0 & 17.495 & 1.252 & 2.7804$\times 10^{-2}$ $\pm$ 1.0220$\times 10^{-4}$ \\
 & 4096 & 0.50 & 5.0 & 18.652 & 2.014 & 4.4709$\times 10^{-2}$ $\pm$ 8.6971$\times 10^{-5}$ \\
 & 8192 & 0.50 & 5.0 & 19.367 & 4.886 & 1.0848$\times 10^{-1}$ $\pm$ 2.1029$\times 10^{-3}$ \\ \cline{1-7} 
\end{tabular}
\caption{Table of proposed parameters \textcolor{black}{(in red asterisk marks in Fig. \ref{Fig.AUCpowersNcost})} of $\epsilon = (\alpha, c)$ for data samples ($N$) under various background noises. The selected parameters provide the maximal averaged AUPC. The relative computational cost is the relative value calculated based on the computational time of each noise with $N=512$. The runtime solely depends on the selected $\alpha, ~N$, and $c$, regardless of the choice of the noise type.}
\label{tab:optparameters}
\end{small}
\end{table*}

\section{Discussions} \label{sec:discussions}

This study investigated various aspects of computing MICe to address the questions posed in the introduction section. These questions are frequently asked while using MICe in the \ac{Gw} data analysis with multi-channel datasets. Moreover, similar questions could emerge in other fields of science and engineering. According to the analysis results provided in this study, we can establish a strategy for setting the parameters required to compute MICe between datasets from multi-channel devices. First, we can check the noise level between two datasets by computing ${\mathcal N}$ expected to be ${\mathcal O}(1)$. If the discord of the noise level is severe, we can perform a relatively denoising procedure to obtain a comparable noise level. Subsequently, we perform the resampling process to maintain a sufficient data sample size. Finally, we compute the averaged AUPC and select a set of MICe parameters, $(\alpha, c)$, with a reasonable computational cost.

As summarized, it is observed that the statistical power of MICe depends on the choice of parameter sets, noise level of data, and data sample size. Furthermore, the values of the parameters rely on the type of background noise and data sample size adopted. To compute some gauges of ${\mathcal N}$ and ${\mathcal P}^{\mathrm{MICe}}$, we can select the set of parameters $(\alpha, c)$, yielding the maximal result of the statistical power. To handle the data of different sampling frequencies, it is crucial to have a sufficient data sample size, regardless of the resampling scenarios selected. Even if we can improve the MICe algorithm by suggesting other methods, it is appropriate to identify the non-linear couplings between two variables from different channels. To make a more reliable decision, it is important to have a consistent standard for interpretations. 

\section*{Acknowledgments}
%\vspace{-2mm}
The authors would like to thank Kyujin Kwak, Kyungmin Kim, and Whansun Kim for their helpful discussions and comments.
This work was supported by the Basic Science Research Program through a National Research Foundation of Korea (NRF) funded by the Ministry of Education (NRF-2020R1I1A2054376). Furthermore, this work is partially supported by the NRF grant funded by the Korean government's Ministry of Science and ICT (No.\ 2019R1A2C2006787, No.\ 2016R1A5A1013277, No.\ 2021R1A2C1093059, and  No.\ 2022R1C1C2012226). 
%\vspace{-5mm}

%\newpage
\appendix

\section{Mutual Information and Maximal Information Coefficient}
\label{sec:appendixA}

MI can be measured by their dependencies on two datasets, defined by the Kullback-Leibler divergence as:
\begin{equation}
    I(X;Y) = D_{KL} (P(X,Y)||P(X)P(Y))
\end{equation}
between the product of two marginal probability $P(X)/P(Y)$ and the joint probability $P(X, Y)$. 
The quantity can be rewritten for the discrete data as:
\begin{equation}
    \label{eq:mi}
    I(X;Y) = \sum_{y\in Y} \sum_{x\in X} p(x,y) \log \left(\frac{p(x,y)}{p(x)p(y)} \right),
\end{equation}
where $p(x,y)$ is the joint probability function of two datasets $X$ and $Y$, and $p(x)/p(y)$ is the marginal probability function of $X/Y$. The mutual information measures the amount of information shared between two datasets. Therefore, when the two datasets have no shared information $p(x,y)=p(x)p(y)$, then $I(X;Y)$ clearly disappears. MI can be defined on the $a \times b$ grid as:
\begin{equation} \label{eq: MI}
I(X, Y)|_{a,b} = \sum_{i=1}^{a}\sum_{j=1}^{b}p_{XY}(i,j) \log_{2}{\frac{p_{XY}(i,j)}{p_{X}(i)p_{Y}(j)}},
\end{equation}
where $p_{XY}(i,j)$ is the joint probability distribution obtained by the occupancy of the elements of the ($i$, $j$)-th bin, as well as the $p_{X}(i)$ and $p_{Y}(j)$ marginal distributions on $i$-th columns and $j$-th rows, respectively. In addition, $a$ and $b$ denote the total number of each grid size. 
Hence, MIC can be defined as:
\begin{equation} \label{eq: MIC}
MIC(X, Y , \alpha, c) = \max_{xy<B({\mathfrak N})}  M_{a,b},
\end{equation}
where $M_{a,b}$ indicates the characteristic matrix that consists of the highest normalized MI in Eq. (\ref{eq: MI}) given by an $a\times b$ grid expressed as:
\begin{equation} \label{eq: ChracteristicMatrix}
M_{a,b} = \Big\{ \max \Big\{ \frac{ I^{*}(X, Y, c)}{\log_2 \min\{a,b\}} \Big\} : ab \le B({\mathfrak N})  \Big\}.
\end{equation}

MIC explores every possible $a\times b$ grid up to maximal bin resolutions and selects the maximal value on them. The association between two datasets can be visualized when the scattered plot is drawn on the $X-Y$ plane.

To control the matrix range in Eq. (\ref{eq: ChracteristicMatrix}), the parameter $\alpha$ is restricted up to the upper bound of bin sizes by $B({\mathfrak N})={\mathfrak N}^{\alpha}$, $0 \le \alpha \le 1$. For instance, if the $B({\mathfrak N})$ obtained is negligible, then the grid patterns become too simple to lose generality; however if it is significantly large, then a non-trivially high score of MIC is obtained for the randomly selected paired samples. For a larger sample size ${\mathfrak N}$, the computational cost is expected to be expensive owing to the search for an optimal value for all possible grids. In addition, the parameter $c$ restricts a maximum number of partitions to reduce the redundant computing power in the maximizing process of MI, called ApproxiMaxMI,
\begin{equation} \label{eq: ApproxiMaxMI}
\begin{array}{l}
I^*(X, Y, c)|_{a,b} = \mathrm{max}\Big\{ \ I(X, Y)|_{l, k}: l \in [2, ca], \ 2 \le k \le l < b  \Big\}.
\end{array}
\end{equation}
Given an $a \times b$ grid resolution, ApproxiMaxMI provides maximum values by finding an optimal grid on every possible partition size,  $l$ and $k$, within the bound limitations. Subsequently, the parameter $c$ determines the complexity of the grid partitioning process.

Reshef {\it et al.} \cite{JMLR:v17:15-308} also proposed an empirical estimator of MIC (MICe), as expressed in Eq. (\ref{eq:mice}). The value of MICe varies from zero to one depending on the strength of association between the two variables. It is unity when sharing information is maximal for sufficiently large data sizes; however, it becomes zero for the null association. For more details, refer to \cite{reshef_detecting_2011, MICe2018, JMLR:v17:15-308, MICe2014}.

 \bibliographystyle{elsarticle-num} 
 \bibliography{MICRef}

\end{document}